\documentclass[conference]{IEEEtran}
\IEEEoverridecommandlockouts
\usepackage{cite}
\usepackage{amsmath,amssymb,amsfonts}
\usepackage{algorithm}
\usepackage{algpseudocode}
\usepackage{graphicx}
\usepackage{textcomp}
\usepackage{xcolor}
\usepackage{fancyhdr}
\usepackage{caption,subcaption}
\usepackage[bookmarks=true,breaklinks=true,letterpaper=true,colorlinks,linkcolor=black,citecolor=blue,urlcolor=black]{hyperref}
\usepackage{listings}
\usepackage[export]{adjustbox}
\usepackage{multirow}

\def\BibTeX{{\rm B\kern-.05em{\sc i\kern-.025em b}\kern-.08em
    T\kern-.1667em\lower.7ex\hbox{E}\kern-.125emX}}

\fancypagestyle{firstpage}{
  \fancyhf{}

  \fancyhead[C]{} 
  \fancyfoot[C]{\thepage}
}  

\title{Per-Bank Bandwidth Regulation of Shared Last-Level Cache for Real-Time Systems
}

\author{
\IEEEauthorblockN{Connor Sullivan}
\IEEEauthorblockA{
\textit{University of Kansas}\\
Lawrence, Kansas USA \\
connor.sullivan13@ku.edu}
\and
\IEEEauthorblockN{Alex Manley}
\IEEEauthorblockA{
\textit{University of Kansas}\\
Lawrence, Kansas USA \\
amanley97@ku.edu}
\and
\IEEEauthorblockN{Mohammad Alian\textsuperscript{\textasteriskcentered}}
\IEEEauthorblockA{\textit{Cornell University}\\
Ithaca, New York USA \\
malian@cornell.edu}
\and
\IEEEauthorblockN{Heechul Yun}
\IEEEauthorblockA{
\textit{University of Kansas}\\
Lawrence, Kansas USA \\
heechul.yun@ku.edu}
}

\begin{document}
\maketitle
\begingroup\renewcommand\thefootnote{\textasteriskcentered}
\footnotetext{This work was conducted while affiliated with the University of Kansas.}
\thispagestyle{firstpage}
\begin{abstract}
Modern commercial-off-the-shelf (COTS) multicore processors have advanced memory hierarchies that enhance memory-level parallelism (MLP), which is crucial for high performance. To support high MLP, shared last-level caches (LLCs) are divided into multiple banks, allowing parallel access. However, uneven distribution of cache requests from the cores, especially when requests from multiple cores are concentrated on a single bank, can result in significant contention affecting all cores that access the cache. Such cache bank contention can even be maliciously induced---known as cache bank-aware denial-of-service (DoS) attacks---in order to jeopardize the system's timing predictability. 

In this paper, we propose a per-bank bandwidth regulation approach for multi-banked shared LLC based multicore real-time systems. By regulating bandwidth on a per-bank basis, the approach aims to prevent unnecessary throttling of cache accesses to non-contended banks, thus improving overall performance (throughput) without compromising isolation benefits of throttling. 
We implement our approach on a RISC-V system-on-chip (SoC) platform using FireSim and evaluate extensively using both synthetic and real-world workloads. Our evaluation results show that the proposed per-bank regulation approach effectively protects real-time tasks from co-running cache bank-aware DoS attacks, and offers up to a 3.66$\times$ performance improvement for the throttled benign best-effort tasks compared to prior bank-oblivious bandwidth throttling approaches. 
\end{abstract}

\pagestyle{plain}
\section{Introduction} \label{sec:intro}

Modern commercial-off-the-shelf (COTS) multicore processors are equipped with sophisticated memory hierarchies that support a high degree of memory-level parallelism (MLP). Because memory accesses often take significantly longer than actual computation, enabling high MLP across all levels of the memory hierarchy is crucial for achieving high performance in modern multicore architectures.

To facilitate high MLP, shared last-level caches (LLCs) are often organized into multiple banks that can be independently accessed in parallel. For instance, the LLC of the ARM Cortex-A72 processor has two independent tag banks, each of which is further divided into four data banks~\cite{a72}. Such a multi-bank cache design maximizes parallelism and throughput in accessing the cache, and is widely adopted in high-performance multicore architectures~\cite{a57,a72,arm-dsu,pi5,beaglev}, including those that are used in safety-critical embedded real-time systems in automotive and aviation domains~\cite{t4080,l2080}.

While most prior work on shared cache for real-time systems has focused on cache space partitioning, multiple studies have shown that partitioning cache space alone does not guarantee temporal isolation in accessing the cache~\cite{valsan2016taming,bechtel2019dos,iorga2020slow,li2022polyrhythm,bechtel2023attack}. In particular, it has been shown that the performance of a multi-bank cache can degrade significantly when requests to the cache are unevenly distributed across the banks. In the worst-case scenario, when all requests are concentrated on a single cache bank, severe contention can arise.  
Such bank conflicts can disrupt the system's temporal predictability and be leveraged as cache bank-aware denial-of-service (DoS) attacks~\cite{bechtel2023attack}. 

To mitigate 
shared cache bank contention, the prior study~\cite{bechtel2023attack} suggested a software-based cache bandwidth throttling approach as a potential solution, which is based on MemGuard~\cite{yun2013memguard} and uses hardware performance counters to monitor and regulate the LLC access bandwidth of the offending cores (those that generate excessive parallel requests to the LLC). However, such a software-based bandwidth throttling solution severely impacts the performance of
the throttled cores. To provide sufficient isolation for the protected real-time task, it reportedly incurs up to 300$\times$ slowdown of the throttled tasks ~\cite{bechtel2023attack}, which may be unacceptable overhead for many applications.
While hardware-based memory bandwidth throttling solutions~\cite{rdt, mpam, farshchi2020bru}, if used for LLC bandwidth throttling, can potentially reduce the overhead of software-based throttling, their effectiveness is still fundamentally limited because they are not aware of cache banks when regulating bandwidth, which makes them overly pessimistic.

In this paper, we propose per-bank bandwidth regulation of shared LLCs for predictable and efficient use of the shared cache in multicore SoCs for real-time systems. Our approach is motivated by the observation that the worst-case bank contention arises when cache accesses are concentrated on a single cache bank rather than distributed across the banks. 
As such, instead of throttling bandwidth to the entire shared LLC, we apply bandwidth throttling on a per cache-bank basis to only throttle accesses when there is a bank conflict. This effectively multiplies the permissible cache access bandwidth of best-effort tasks without compromising the isolation benefits of bandwidth throttling to the protected real-time tasks.

We implement the proposed per-bank throttling capability as an extension to an open-source hardware memory bandwidth regulator~\cite{farshchi2020bru} on a RISC-V system-on-chip (SoC) platform using Xilinx UltraScale+ VCU118 FPGA~\cite{vcu118} and FireSim~\cite{karandikar2018firesim}. 
We evaluate the effectiveness of the proposed approach in providing temporal isolation to the real-time victim tasks in the presence of cache bank-aware DoS attacks. We then demonstrate the efficiency benefits of per-bank regulation over prior approaches that throttle the aggregate bandwidth of all banks globally. We show that per-bank regulation can effectively protect victim tasks from the attack while providing best-effort tasks with up to a 3.66$\times$ performance improvement over the prior bank-oblivious regulation scheme.

In summary, we make the following contributions:
\begin{itemize}
	\item We propose per-bank bandwidth regulation on shared LLC to effectively and efficiently defend against potential cache bank contention attacks (regardless of whether malicious or benign).
	\item We present a prototype hardware design, which can be integrated into any RISC-V SoC that supports the standard TileLink interconnect, and analyze its ability to prevent cache bank-aware DoS attacks. 
	\item We implement our design on a realistic cycle-exact, FPGA-accelerated full-system simulator, and evaluate its performance improvements over prior bank-oblivious regulation approaches. 
    We also provide our design as open-source\footnote{\url{https://github.com/CSL-KU/per-bank-regulation-firesim}}. 
\end{itemize}

The remainder of the paper is organized as follows. Section~\ref{sec:background} provides the necessary background. 
Section~\ref{sec:threatmodel} defines the threat model. 
Section~\ref{motivation} motivates the need for per-bank regulation. We present our proposed per-bank regulation design in Section~\ref{sec:solution} and the evaluation results in Section~\ref{eval}. We discuss related work in Section~\ref{related} and conclude in Section~\ref{conclude}.

\section{Background} \label{sec:background}

In this section we provide the necessary background on multi-banked caches, cache bank-aware DoS attacks, and bandwidth regulation methods.

\subsection{Multi-Bank Cache Organization}
The shared cache of a modern multicore processor is often composed of multiple independent banks (sometimes referred to as slices~\cite{cachehierarchy}), which can be accessed in parallel.
This multi-bank cache organization facilitates high MLP, 
which is crucial for high-performance multicore processors.  
In a multi-bank cache, a mapping function determines the bank from a given physical address. The mapping function can be as simple as using a subset of the memory address bits.

\begin{figure} [htp]
    \centering
    \includegraphics[width=0.5\textwidth]{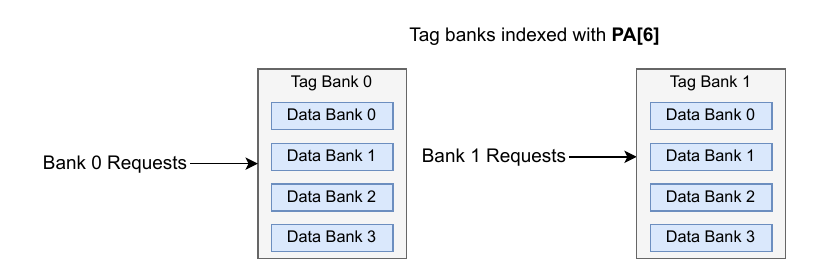}
    \caption{ARM Cortex-A72 LLC Organization~\cite{a72}. }
    \label{fig:a72-cache}
\end{figure}

Figure~\ref{fig:a72-cache} depicts the multi-bank LLC organization of the ARM Cortex-A72~\cite{a72}. Note that it is comprised of two independent tag banks, each of which is further divided into four sub-banks called data banks. The tag banks are completely independent, allowing for two separate LLC accesses to be serviced in parallel. Likewise, the data banks allow for further interleaving of accesses. To index the tag and data banks, physical address bits 4, 5 and 6 are used. Bit 6 indexes between the two tag banks, with bits 4 and 5 being used to index the data banks within each tag bank. For 64 byte cache lines, each line is split into four sub-lines of 16 bytes that are striped across the data banks. 

It is important to note that these bits (4, 5 and 6) are in the lower 12 bits of an address, within the page offset. This means that these bits can be fully controlled from the user space without the need for elevated privileges or huge pages. With this understanding, an attacker can direct memory accesses to specific banks, opening the door for potential DoS attacks~\cite{bechtel2023attack}.

\subsection{Cache Bank-Aware DoS Attack}
The feasibility of cache bank-aware DoS attacks was first demonstrated in a recent study~\cite{bechtel2023attack} on both ARM Cortex-A57~\cite{a57} and Cortex-A72~\cite{a72} cores. Under the threat model described in Section~\ref{sec:threatmodel}, the study shows that by saturating a single cache bank of the shared L2 cache with many parallel requests, an attacker can cause up to a 10$\times$ cross-core slowdown on a victim task. This slowdown occurs even when the victim is running on a dedicated core in isolation, accessing a dedicated L2 cache (space) partition by means of page coloring. 
As the cache space is partitioned between the victim and the attacker,
it demonstrates that the slowdown is not caused by cache evictions. 
Furthermore, it also shows that the contention occurs at the bank level---not at the bus level---as no slowdown is observed when the victim and attacker target separate cache banks. Lastly, the worst-case slowdown occurs when both the victim and the attacker access the same cache bank, suggesting that the cache bank bandwidth becomes the bottleneck in such a situation. 

\subsection{Cache Bandwidth Regulation}

To mitigate cache bank-aware DoS attacks, the prior study~\cite{bechtel2023attack} proposed a software-based cache bandwidth regulation method, LLCGuard, which uses per-core performance counters to regulate (limit) each core's LLC access bandwidth (as opposed to DRAM bandwidth regulation proposed by MemGuard~\cite{yun2013memguard}) at a regular time interval (e.g., 1ms). However, the software-based approach is known to incur very high performance cost to the throttled best-effort cores. Concretely, the study reports up-to 300$\times$ slowdown of the tasks on the throttled cores to ensure no more than 1.1$\times$ slowdown of the protected real-time tasks on the unregulated core. 
As discussed in~\cite{bechtel2023attack}, part of the reason for such a massive performance loss is due to software implementation overhead. With the regulation period of 1ms, a large amount of LLC accesses can still occur in short bursts, which results in LLC bank contention.

In contrast, a hardware-based cache bandwidth solution can operate at a much finer granularity (in cycles), which can help spread the LLC accesses more evenly across the entire throttle period, thereby reducing the negative performance impact of throttling best-effort cores. While existing hardware-based bandwidth regulators, such as
Intel RDT~\cite{rdt} and ARM MPAM~\cite{mpam}, are mainly designed to regulate memory bandwidth, they can potentially be modified to regulate cache bandwidth to mitigate cache bank contention. 

Unfortunately, all aforementioned regulation schemes, both software and hardware, suffer from a common limitation---they do not regulate at the \textit{bank level}, where the actual contention occurs. Instead, they treat the entire cache (or DRAM) as a single resource and regulate its total access bandwidth. We henceforth refer to the latter as \textit{all-bank} regulation. In the following, we show why this all-bank regulation is overly \textit{pessimistic}. 

\section{Threat Model of Cache Bank DoS Attacks} \label{sec:threatmodel}
In this work, we consider the same threat model used in~\cite{bechtel2023attack}.
That is, we assume: (1) a victim task and one or more attacker
tasks are co-located on a multicore platform, which has a shared last-level cache (LLC) and main memory (DRAM); (2) the victim and the attackers are partitioned to run on dedicated CPU cores and LLC cache spaces; (3) the attackers have non-privileged access on the target platforms and can only execute code
from the userspace; (4) the cache bank address mapping information is known beforehand either from datasheets~\cite{a57, a72} or reverse engineering~\cite{farshin2019make,irazoqui2015systematic}. 
Following these conditions, our goal is to guarantee temporal isolation of the victim accessing the shared cache in the presence of co-scheduled attackers, while maximizing cache bandwidth throughput available to the attackers. 

\section{Motivation} \label{motivation}

In this section, we first evaluate the effect of cache bank-aware DoS attacks, synthetic workloads that generate severe cache bank contention\cite{bechtel2023attack}, on two embedded multicore platforms (Section~\ref{sec:realplatforms}). We then discuss the limitations of  bank-oblivious ``all-bank'' cache bandwidth regulation approaches in mitigating such attacks (Section~\ref{sec:allbanklimits}). 

\subsection{Effects of Cache Bank-Aware DoS Attacks}\label{sec:realplatforms}
In this experiment, we use two contemporary embedded multicore platforms: Raspberry Pi 4 Model B~\cite{pi4} and BeagleV Ahead~\cite{beaglev}. The Raspberry Pi 4 is based on the Broadcom BCM2711 SoC and is equipped with four ARM Cortex-A72~\cite{a72} cores with a 1MB shared L2 cache. Comparably, the BeagleV Ahead is based on the Alibaba T-Head TH1520 SoC, equipped with four Xuantie C910 RISC-V cores with a 1MB shared L2 cache. Table~\ref{tbl:platforms} shows the basic characteristics of the two platforms. 

\begin{table}[htp]
  \centering

  \begin{tabular}{|c||c|c|}
    \hline
    Platform                & Raspberry Pi 4 (B) & Beagle V Ahead \\ 
    \hline
    SoC                     & BCM2711           & TH1520         \\ 
    \hline
    Architecture            & ARMv8-A            & RISC-V 64GC \\
    \hline
    \multirow{4}{*}{CPU}    & 4x Cortex-A72  & 4x Xuantie C910   \\ 
                            & out-of-order  & out-of-order          \\ 
                            & 1.5GHz        & 2.0GHz               \\ 
                            & 48KB(I)/32KB(D)   & 64KB(I)/64KB(D)   \\
    \hline
    Shared L2 Cache       & 1MB       & 1MB       \\ 
    \hline 
    Memory                  & 4GB LPDDR4          & 4GB LPDDR4             \\
    \hline
  \end{tabular}
  \caption{COTS embedded multicore platforms.}
  \label{tbl:platforms}
\end{table}

For software, the Pi 4 runs Raspberry Pi OS with Linux kernel 6.6, and the Beagle V runs Ubuntu 20.20 with Linux kernel 5.10. In both platforms, the kernels are patched with PALLOC~\cite{yun2014palloc}, a page coloring mechanism for Linux, to partition the L2 cache space equally between the victim and the attackers. 

For evaluation, we use the \textit{BkPLL} workload from~\cite{bechtel2023attack}. As both the victim and the attackers, \textit{BkPLL} is a pointer chasing workload that can generate a configurable number of parallel memory requests targeting a specific cache bank. We first run the victim on one core in isolation and measure its performance. We then repeat the experiment in the presence of co-running attacker tasks on the other cores. 
We evaluate different combinations of target cache banks for the victim and the attackers: \emph{Same Bank} refers to the case where the victim and the attacker target the same cache bank, whereas \emph{Diff Bank} refers to the case where they target different banks. 

\begin{figure} [htp]
    \centering
    \includegraphics[width=0.5\textwidth]{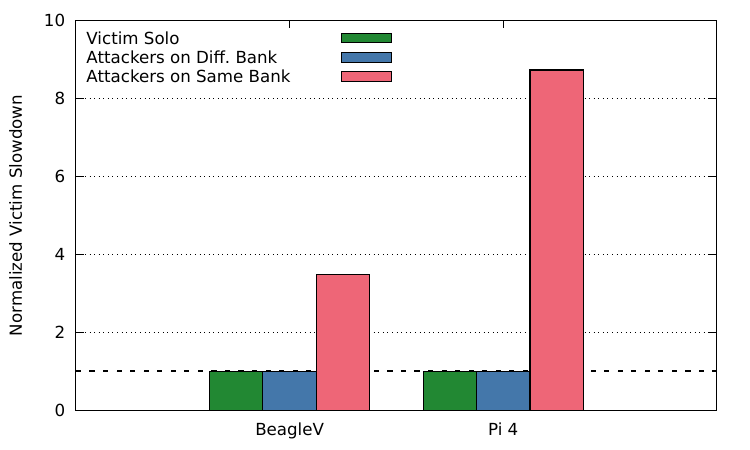}
    \caption{Effects of cache bank-aware DoS attacks}
    \label{fig:dev-boards}
\end{figure}

Figure~\ref{fig:dev-boards} shows the normalized slowdowns of the victim on different cache bank mapping configurations. The dashed horizontal line denotes the baseline 1.00$\times$ slowdown (in this case, solo performance). First we notice that, consistent with the findings in~\cite{bechtel2023attack}, contention occurs at the cache bank level, not on the shared bus level. The victim sees no slowdown when the attackers target a different bank, whereas severe slowdown is observed when both target the same cache bank. Second, we observe up to 8.7$\times$ slowdown on the Pi 4 platform, which is considerably worse than the 8.3$\times$ reported worst-case slowdown on the same platform~\cite{bechtel2023attack}. Interestingly, we find different target data bank selections for the victim and the attackers contribute to the increased worst-case slowdown. Third, the BeagleV platform shows similar trends but its worst-case slowdown is considerably less (3.5$\times$) than that of the Pi 4 (8.7$\times$). 
This is due to the differences in baseline performance---i.e., the CPU core's ability to concurrently generate requests and the peak bandwidth of the cache. Note that Raspberry Pi 4's peak cache bandwidth is 2$\times$ higher than that of the BeagleV. In general, faster processors tend to suffer larger worst-case slowdowns.

\subsection{Limitations of ``All-Bank'' Bandwidth Regulation}~\label{sec:allbanklimits}
The results in the previous subsection show that the contention created by the DoS attack is not on the shared bus, but at the targeted cache bank. This indicates that, in order to mitigate the bank contention attack, we only need to limit (throttle) the traffic (bandwidth) going into the contended bank. Furthermore, the banks in the cache are independent of each other. As such, regulation should be applied on a per-bank basis rather than applied unnecessarily across all banks. Unfortunately, existing bandwidth regulation approaches cannot be applied to individual cache banks.

For example, BRU~\cite{farshchi2020bru} is a hardware-level bandwidth regulator inserted between the L1 caches and the shared L2 cache~\cite{farshchi2020bru}. As such, it regulates the L2 access traffic of the subset of cores that may be executing the DoS attackers. However, BRU tracks all L2 access traffic, 
without consideration for the individual bank destination. In other words, it  implements an ``all-bank'' bandwidth regulation scheme.

\begin{figure}[htp]
    \centering
    \includegraphics[width=0.45\textwidth]{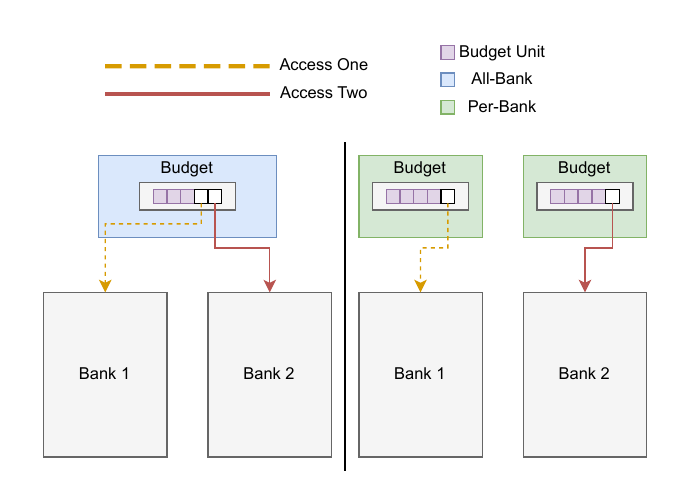}
    \caption{All-bank vs. per-bank bandwidth regulation on multi-bank shared caches}
    \label{fig:motivation}
\end{figure}

Figure~\ref{fig:motivation} depicts the high-level intuition illustrating why all-bank regulation is needlessly pessimistic. Consider two bandwidth regulation systems, one with all-bank regulation (left) and one with per-bank (right). Both systems have two cache banks. 
Suppose that we need to limit the traffic to 5 accesses to a cache bank per regulation period to mitigate the contention on the bank. 
In the case of all-bank regulation it is bank oblivious, thus the global cache access budget must be set to 5 accesses, to counter the worst-case where all traffic goes to one cache bank. This budget is deducted on every cache access, even though in reality, the two cache accesses are interleaved across two different banks. 

Per-bank regulation, in contrast, can apply the access budget of 5 to each bank separately, only then is a budget deducted when the specific bank is accessed. As such, when the two accesses are interleaved over the two banks, each bank's budget is depleted by one, leaving a remaining budget of 4 for each bank, whereas only 3 would be left in the all-bank case. 
As more accesses are interleaved, per-bank regulation can provide higher aggregate bandwidth while still providing worst-case cache bank contention guarantees. 
With this intuition in mind, we now discuss our proposed per-bank cache bandwidth regulation system design.

\section{Per-Bank Cache Bandwidth Regulation} \label{sec:solution}
In this section, we describe the design and implementation details of the proposed per-bank cache bandwidth regulation approach. 

\subsection{Design Overview} \label{sec:design}

Our per-bank bandwidth regulation solution is implemented as an extension to an open-source hardware bandwidth regulator called BRU~\cite{farshchi2020bru}, designed to drop into an SoC design 
between the cores/accelerators and the shared cache. Figure~\ref{fig:high-level-design} depicts a high-level view of our bandwidth regulation unit in a basic dual-core setup.

\begin{figure}[htp]
    \centering
    \includegraphics[width=0.28\textwidth]{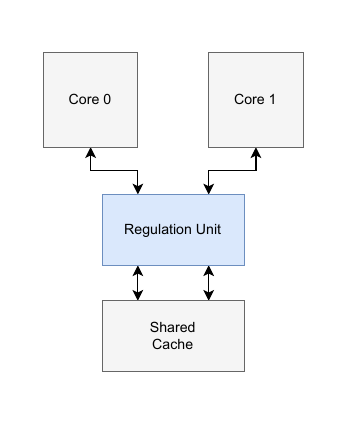}
    \caption{High-level view of a regulation unit in a dual-core SoC}
    \label{fig:high-level-design}
\end{figure}

BRU supports creation of multiple arbitrary domains, each of which may be composed of one or more cores. 
A domain is the primary entity that bandwidth regulation is applied to. In the original BRU design, each domain can be configured with a period (cycles) and a budget (number of memory requests). The budget is decremented for any memory request made to the shared cache (regardless of which bank it targets) and once the budget is depleted, all cores in the domain are denied access to the shared cache until the period expires and the budget is replenished. As discussed earlier, we call this all-bank regulation because it counts an access to any bank equally.

Our modifications to BRU enable tracking and regulating the budget for each shared cache bank rather than for the entire cache. Specifically, for each request to the cache, we decode its destination cache bank address and charge it to the corresponding bank's budget. This means, for an $N$ bank shared cache, we have $N$ separate bank budgets to keep track of. When any one bank's budget is depleted, further access to the bank will be prevented(throttled) until the next period begins. Accesses to other banks can still occur as long as their budgets are not depleted. 

\subsection{Per-bank Bandwidth Regulation Interface}
To enable fixed user-defined bandwidth regulation, our design exposes memory-mapped I/O (MMIO) registers. The \textit{Access Budget Register (ABR)} is used to program the maximum number of accesses per period and the \textit{Regulation Period Register (RPR)} sets the regulation period in cycles. Equation~\ref{eq:bandwidth} represents the bandwidth budget assigned to each bank given the values of the \textit{ABR} and \textit{RPR} registers. Note that each bank gets the budget \textit{BW}, rather than it being evenly distributed among the banks. \textit{TS} represents the \textit{transaction size} and \textit{f} the \textit{clock frequency}. When a core accesses the cache, \textit{TS} is equivalent to the line size (commonly 64 bytes).

\begin{equation}
\text{BW} =  \frac{\text{\textit{ABR}}}{\text{\textit{RPR}}} \times \text{\textit{TS}} \times f
\label{eq:bandwidth}
\end{equation}

The \textit{RPR} is applied globally to all logical groupings of cores that are being regulated concurrently---we refer to these groupings as regulation \textit{domains}. 
Each domain has an \textit{ABR}, allowing unique per-domain budgets to be applied to each cache bank.

Along with a budget configuration interface, fixed bandwidth regulation requires mechanisms to track accesses (per-bank counters in our design) and regulate these accesses as necessary. We organize the per-bank access counters in a \textit{Domain Control Interface (DCI)}, with a \textit{Core Control Interface (CCI)} containing regulation enable registers and domain assignment registers.

\textbf{Domain Control Interface.} Each domain has its own access counter registers. We denote these registers as \textit{Bank Access Counters (BAC)}. A given domain has \textit{N} of these registers, where \textit{N} is equal to the number of banks in the shared cache. These registers are used solely for bandwidth regulation. Note that this interface also includes the user configurable per-domain \textit{ABR} registers.

\textbf{Core Control Interface.} This interface includes logic for assigning cores to domains and enabling regulation for a given core. Domain assignment is handled through the \textit{Domain Assignment Registers (DAR)}, enabling each core to be configured to any one domain. A \textit{Regulation Enable Register (RER)} is generated for each core and allows for regulation to be enabled or disabled seamlessly.

\begin{figure} [htp]
    \centering
    \includegraphics[width=0.5\textwidth]{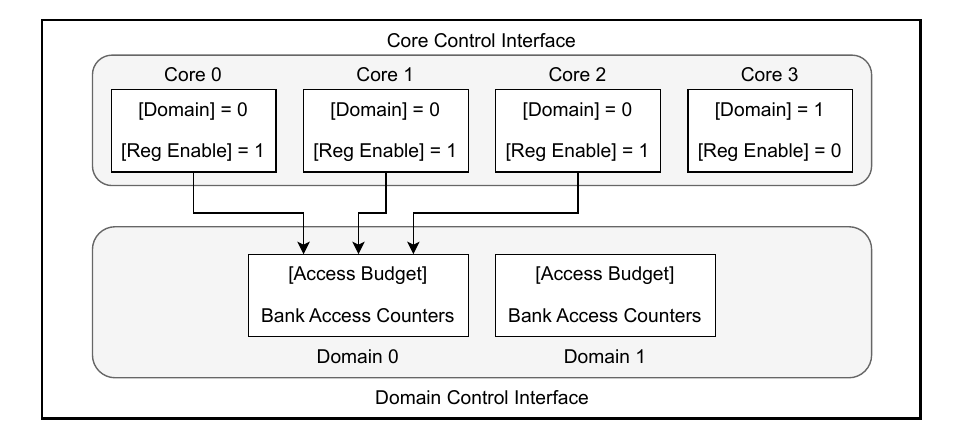}
    \caption{Example of group regulation in a quad-core system. Core 0-2 belong to Domain 0, which is regulated. Core 3, on the other hand, belongs to Domain 1, which is not regulated. In this example, Domain 0 is for \textit{best-effort} tasks while Domain 1 is for the \textit{real-time} tasks.}
    \label{fig:ctrl-interface}
\end{figure}

Figure~\ref{fig:ctrl-interface} depicts the regulation unit's control interface for an arbitrary quad-core system configured to have two regulation domains. In the \textit{CCI} there is logic for the four cores. Cores 0-2 are assigned to Domain 0, with their \textit{RERs} set high. Core 3 is assigned to Domain 1, but its \textit{RER} is kept low, meaning that the core's accesses are not being regulated. Bracketed register names(i.e. [Domain]) indicate that they are memory mapped and user configurable.

\subsection{Per-bank Bandwidth Monitoring Interface}
In addition to per-bank regulation, we also provide a per-bank bandwidth monitoring interface to enable software(OS) based fine-grained monitoring and adaptive bandwidth regulation capabilities. Specifically, our bandwidth regulation unit includes per-bank monitoring registers that are separate from the regulation interface. The per-bank monitoring registers form the \textit{monitoring interface}. For every core in the system, a set of \textit{N} counters is generated, where \textit{N} is equal to the number of banks. Similar to typical performance counters, these counters can be reset and read by a user to determine the per-bank bandwidth and access pattern of a core. 

\subsection{Regulation and Monitoring Algorithms}
Algorithm~\ref{alg:bank-reg} shows the pseudo-code of our regulation unit. At a high level, it manages the global period counter (lines 1-8), the per-bank access counters, and throttling (lines 9-22). 

Each clock cycle, \textit{PeriodCounter} is incremented to advance the current regulation period (line 7). The \textit{PeriodCounter} is reset when its value equals or exceeds the user defined \textit{RegulationPeriod}. As part of this reset, all bank budgets are replenished by zeroing the \textit{BankCounters} (lines 1-6).

Lines 9-22 make up the main body of our regulation algorithm, with the logic being evaluated per-core and per-bank (lines 9 and 10). In a given domain, if the budget of bank $j$ is depleted, then all further accesses to that bank are stalled for the cores in that domain (lines 13-16). When a core sends a bank access, the corresponding domain's bank access counter is incremented (lines 17-20). Specifically, bank accesses occur on Channel A ($A(i).isAccess$), a TileLink notation which we further explain below (Section~\ref{sec:imp}).

\begin{algorithm}
\caption{Per-bank Regulation Algorithm}
\label{alg:bank-reg}
\begin{algorithmic}[1]
\If {PeriodCounter $\geq$ RegulationPeriod}
    \State PeriodCounter $\gets$ 0
    \ForAll {c in BankCounters}
        \State c $\gets$ 0 
    \EndFor
\Else
    \State PeriodCounter++
\EndIf

\For {i $\gets$ 0 to nCores - 1}
    \For {j $\gets$ 0 to nBanks - 1}
        \State stall(i)(j) $\gets$ False
        \State AccessIsBank(i)(j) $\gets$ (j == bankBits)
        
        \If {(BankCounters(Domain(i))(j)$\geq$AccessBudget) \\
            \hspace{20mm} \textbf{and} AccessIsBank(i)(j)}
            \State stall(i)(j) $\gets$ True
        \EndIf

        \If {A(i).isAccess \textbf{and} AccessIsBank(i)(j)}
            \State BankCounters(Domain(i))(j)++
        \EndIf
    \EndFor
\EndFor
\end{algorithmic}
\end{algorithm}

The monitoring interface is handled similarly. Algorithm~\ref{alg:bank-monitor} shows the pseudo-code for the monitoring interface. The logic happens per-core and per-bank (line 1 and 2). Lines 3-8 are similar to the main body of~\ref{alg:bank-reg}, where a bank monitor counter is incremented when that specific bank is accessed. The difference being that there is no notion of domains or period in the monitoring interface.

\begin{algorithm}
\caption{Per-bank Monitoring Algorithm}
\label{alg:bank-monitor}
\begin{algorithmic}[1]
\For {i $\gets$ 0 to nCores - 1}
    \For {j $\gets$ 0 to nBanks - 1}
        \State AccessIsBank(i)(j) $\gets$ (j == bankBits)
        
        \If {CoreAccess(i) \textbf{and} AccessIsBank(i)(j)}
            \State BankMonitor(i)(j)++
        \EndIf
    \EndFor
\EndFor
\end{algorithmic}
\end{algorithm}

Note that our implementation is written in the Chisel hardware description language (HDL)~\cite{bachrach2012chisel}. This allows our design to support any number of domains and cache banks through configurable parameters, eliminating the need to modify the hardware design code.

\subsection{Implementation} \label{sec:imp}
We implement our design using the Chipyard SoC Framework~\cite{amid2020chipyard}. In this subsection, we discuss details of both the TileLink~\cite{tilelink} interconnect specification and the Rocket Chip SoC~\cite{krste2016rocket} as they relate to our implementation.

Our regulation unit interfaces with TileLink Cached (TL-C) edges. TL-C edges connect the cores to the shared memory subsystem and are cache coherent~\cite{tilelink}. There are five channels of communication on TL-C edges: \textit{A, B, C, D }and \textit{E}. We focus on \textit{Channel A}, which carries requests from the core's private caches to the shared caches and memories.

\begin{figure}[htp]
    \centering
    \includegraphics[width=0.45\textwidth]{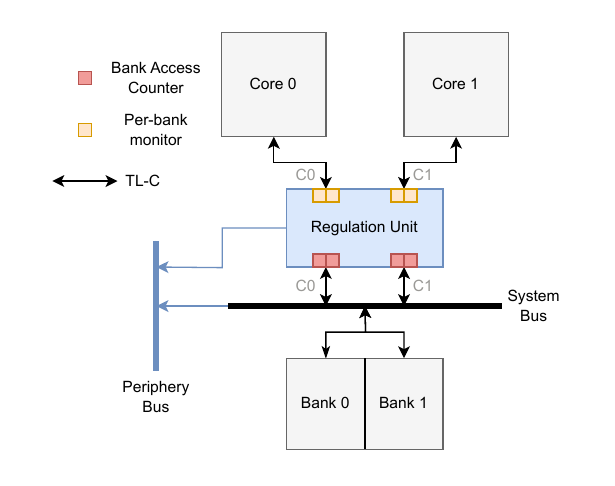}
    \caption{Dual-core Rocket SoC with per-bank regulation unit}
    \label{fig:design}
\end{figure}

Figure~\ref{fig:design} depicts our bandwidth regulation unit in a generic dual-core Rocket Chip SoC design. The connections between the cores and the shared system bus are TL-C edges. When a data or instruction cache miss occurs in the private L1 caches of the cores, a request is sent over Channel A. By monitoring this channel we can track per-core accesses and regulate when a domain's bank budget is depleted.

For synchronizing messages on a given channel, TileLink uses a ready-valid interface for sender/receiver handshaking. To regulate a channel, we can simply set the ready and valid signals to low, effectively stalling the request. Channel A also carries information about the the memory address being requested. From this we can extract the bank address bits to count per-bank accesses. All TL-C channels going from the core to the shared system bus pass through the regulation unit. However, only Channel A is monitored for accesses and regulated via the ready-valid signals. All other channels and signals remain unmodified, passing through and connecting directly to the shared system bus.

Along with connections to the system bus, the regulation unit also connects to the \textit{periphery bus}. This bus is utilized by cores to read/write to the MMIO registers. Figure~\ref{fig:design} also depicts the \textit{BAC} counter registers and monitoring interface registers.

\section{Evaluation} \label{eval}

In this section, we evaluate hardware bandwidth regulation's ability to defend against the cache-bank DoS attack and show the benefits of per-bank over all-bank bandwidth regulation.

\subsection{Experimental Setup}\label{sec:setup}
We use FireSim,
an FPGA accelerated cycle-exact full system simulator~\cite{karandikar2018firesim}. 
This allows us to accurately evaluate the performance of the proposed hardware design when deployed in ASIC, which operates at a higher clock (e.g., $>$1GHz) while being simulated on a FPGA at an actual clock speed of 100MHz.

\begin{table}[htpb]
	\centering
	\begin{tabular}{|c||p{5.5cm}|}
    	\hline
    	\multirow{4}{*}{Cores}  & 1$\times$BOOM, 1GHz, out-of-order, 3-wide, ROB: 96, LSQ: 24/24, L1: 32K(I)/32K(D) \\
        & 2$\times$Rocket, 1GHz, in-order, L1: 16K(I)/16K(D), attached with Mempress traffic generators\\
    	\hline
    	Shared L2 Cache & 1MB (16-way) \\
    	\hline
    	Memory & 4GB DDR3 \\
    	\hline
	\end{tabular}
	\caption{Evaluation platform specifications}
	\label{tab:fsim-specs}
\end{table}

Table~\ref{tab:fsim-specs} shows the basic characteristics of the tri-core heterogeneous SoC we constructed on FireSim for evaluation. The SoC is composed of one out-of-order core, the Berkeley Out-of-Order Machine (BOOM)~\cite{asanovi2015boom}, and two in-order Rocket~\cite{krste2016rocket} cores, which are connected to Mempress traffic generators~\cite{mempress}. All cores share a 1MB L2 cache and a 4GB DDR3 main memory subsystem. 

Note that cache bank-aware DoS attacks~\cite{bechtel2023attack} require out-of-order CPU cores to be able to generate many concurrent memory requests on a specific target cache bank. As such, we initially attempted to construct a quad-core BOOM based SoC on FireSim using the \textit{LargeBoom} configuration. However, due to physical constraints of our FPGA platform, we were unable to fit four large BOOM cores simultaneously in the FPGA. 
Furthermore, there is an unresolved bug in BOOM that results in a simulation hang when executing certain memory intensive workloads on multi-core  configurations\footnote{\url{https://github.com/riscv-boom/riscv-boom/issues/690}}. 
As such, in our simulation setup, we instead utilize the Mempress traffic generator~\cite{mempress} to act as the cache bank attacker tasks. 

Mempress is a configurable hardware unit that can generate multiple parallel streams of requests to the shared memory at varying access patterns. Implemented as an on-chip RoCC accelerator~\cite{rocc}, Mempress has access to the shared memory subsystem. For all following experiments, the attackers will be two separate Mempress units targeting the same last-level cache (LLC) bank. 

The Mempress traffic generators are attached to two Rocket cores (one per core). Since Rocket cores are in-order, they cannot create significant contention in the shared cache on their own. However, Mempress enhances the cores by enabling them to generate parallel accesses through the traffic generators, all while still meeting FPGA space constraints. All targets are clocked at 1GHz.

For the shared L2 cache, we use  SiFive's open-source inclusive cache~\cite{inclusive-L2}, which is a real synthesizable hardware cache design that supports a configurable number of cache banks. The bank mapping bits start at address bit 6 for a two bank design, while bits 6 and 7 are used in a four bank design. Throughout our experimentation, we vary the number of LLC banks to be either two or four, but the size and associativity remain constant. Cache lines are set to be 64 bytes.

All simulations are run with the RISC-V version of Linux kernel 6.2. 
For synthetic workloads, we use \textit{BkPLL} (described in section~\ref{sec:realplatforms}) and \textit{Bandwidth} from~\cite{valsan2016taming}. \textit{Bandwidth} accesses a chunk of memory sequentially, striding at a step size of a cache line. Both workloads can be configured to perform either read or write accesses. 
Lastly, the San Diego Vision Benchmark Suite (SD-VBS)~\cite{sdvbs} with CIF input format is used for real-world evaluation.

To ensure all slowdowns are solely due to bank contention and not impacted by set conflict misses, we apply the PALLOC patch to the Linux kernel~\cite{yun2014palloc} to enable cache set partitioning. Using PALLOC, we create two partitions dividing the LLC of 1MB into equal segments of 512KB each. We assign one partition to victim tasks and one partition to best-effort (attacker) tasks.

\subsection{SD-VBS Profiling} \label{sec:sdvbs-profile}
To guide our evaluation, we first profile the workloads from SD-VBS to find each workload's LLC bandwidth and bank access pattern. With our implemented per-bank monitoring interface, we collect these results on a single-core BOOM system with a four bank LLC. It should be noted that we exclude \textit{multi\_ncut} due to long simulation times.

\begin{table} [htp]
    \begin{tabular}{ p{1.5cm}||p{2cm}|p{2cm}  }
         Workload& LLC Read B/W & LLC Write B/W\\
         \hline
         \textbf{Disparity}&\textbf{2663.1}&\textbf{1330.6}\\
         \textbf{MSER}&\textbf{967.9}&\textbf{270.7}\\
         Sift&356.9&90.6\\
         \textbf{Stitch}&\textbf{795.1}&\textbf{405.1}\\
         Localization&55.9&0.326\\
         Tracking&405.8&173.1\\
         SVM&179.6&45.5\\
    \end{tabular}
    \centering
    \caption{SD-VBS LLC bandwidth characteristics (MB/s)}
    \label{tab:sdvbs-stats}
\end{table}

Table~\ref{tab:sdvbs-stats} shows the collected bandwidth results. From this, we select \textit{Disparity}, \textit{MSER} and \textit{Stitch} for best-effort task evaluation in section~\ref{real-world}, as workloads that do not make frequent accesses to the LLC will not be noticeably affected by regulation. Through experimentation, we determine 700MB/s to be a suitable bandwidth threshold.

\begin{table} [htp]
    \begin{tabular}{ p{1.5cm}||p{1cm}|p{1cm}|p{1cm}|p{1cm}  }
         Workload& Bank 1 & Bank 2 & Bank 3 & Bank 4\\
         \hline
         Disparity&5723148&5694269&5679896&5693761\\
         MSER&476464&467716&466571&468930\\
         Sift&1938519&1908043&1868291&1922350\\
         Stitch&3867787&3821181&3786533&3896458\\
         \textbf{Localization}&\textbf{309455}&\textbf{1843}&\textbf{1647}&\textbf{1795}\\
         \textbf{Tracking}&\textbf{496624}&\textbf{252594}&\textbf{251577}&\textbf{259686}\\
         SVM&540773&505279&557619&525794\\
    \end{tabular}
    \centering
    \caption{SD-VBS per-bank LLC access counts}
    \label{tab:sdvbs-bank-count}
\end{table}

Table~\ref{tab:sdvbs-bank-count} shows the per-bank access counts of the SD-VBS workloads across the four cache banks. In our simulated design, a four-bank LLC uses address bits 6 and 7 to index the banks. Of the seven workloads, \textit{Disparity, MSER, Sift} and \textit{Stitch} have an even access spread across all banks. On the other hand, \textit{Localization} and \textit{Tracking} have heavier traffic to specific banks. \textit{Localization} specifically sees 58$\times$ more accesses directed at bank one than the other three banks combined and 187$\times$ more accesses than the least accessed bank (bank two). \textit{Tracking} sees 2$\times$ more accesses directed at bank one than the least accessed bank. These results show that, in most benign (not malicious) workloads, requests to the shared cache are generally distributed evenly across the cache banks, although there are some notable exceptions.

As such, if we use the all-bank regulation approach to defend against potential cache-bank DoS attacks, which target only one bank, we significantly under utilize the cache bandwidth when benign workloads are executed on the throttled best-effort cores, as we will show later in this section.

\subsection{Cache Bank-Aware DoS Attack on FireSim} \label{sec:attack}

To begin our experimentation, we first mount the cache bank-aware DoS attack~\cite{bechtel2023attack} in our simulation environment, establishing the maximum base-line slowdown for our setup. These results are collected on a system with two banks in the LLC.

The experiments are set up as follows. We utilize the \textit{BkPLL} workload described in Section~\ref{sec:realplatforms} as our victim task. The victim is configured to perform reads (denoted as \textit{BkPLLRead}), has a working-set-size (WSS) of 128KB and is executed on the BOOM core. The Mempress attackers are configured to each have a WSS of 64KB. We first run the victim in isolation and measure its performance. 
The attackers are then co-run with the victim in order to see the attackers impact on the victim’s performance. The attackers are applied in the two following scenarios:

\begin{enumerate}
    \item \textbf{Diff. Bank:} The attackers and the victim target different (disjoint) cache banks (victim: bank 0, attacker: bank 1)
    \item \textbf{Same Bank:} Both the attackers and the victim target the same cache bank (both attacker and victim: bank 0).  
\end{enumerate}

\begin{figure} [htp]
    \centering
    \includegraphics[width=0.5\textwidth]{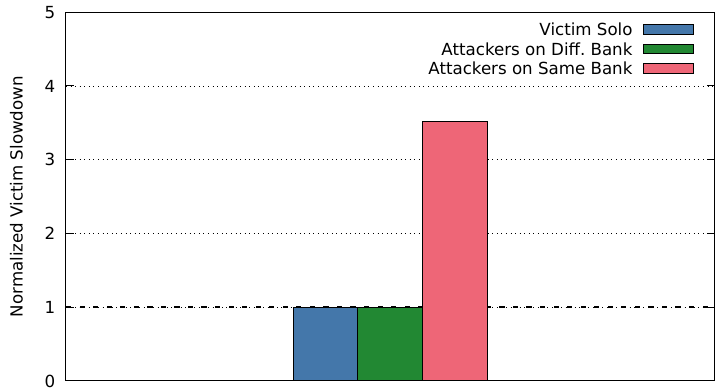}
    \caption{Impact of cache bank-aware DoS attack on synthetic read victim in the FireSim platform. The bank attackers target is varied.}
    \label{fig:bankisol}
\end{figure}

Figure~\ref{fig:bankisol} shows the results. Note first that, when the attackers and victim target separate banks, the victim experiences no slowdown. Yet, when the same bank is targeted the victim experiences a 3.52$\times$ slowdown from the write attackers. 
The results are very similar to what we have observed on the BeagleV platform in Section~\ref{sec:realplatforms}, demonstrating the validity of our evaluation setup. 
To parallel the conclusion on the real platforms, we observe complete temporal isolation when the victim and the attackers target different banks. This confirms that any contention created by the attacker is at the \textit{bank level}, not the interconnect (bus) level. 

Three key takeaways are: (1) each cache bank should be considered as an independent shared resource, which has limited bandwidth. If the bandwidth is over-saturated, then contention occurs and the subsequent requests to the bank will be delayed; (2) the targeted bank DoS attack is effective because it saturates the bank's limited bandwidth; (3) Using bandwidth regulation to mitigate contention by preventing bandwidth saturation will be effective. 

\subsection{Evaluation of Hardware Bandwidth Regulation} \label{sec:threshold}

In this experiment, we evaluate the effectiveness of BRU's bandwidth regulation in providing temporal isolation to the victim task in the presence of cache bank DoS attackers. 

For this experiment, we use all-bank regulation as  implemented in the baseline BRU~\cite{farshchi2020bru}. As discussed earlier, BRU allows for cores to be regulated alone or in groups using \emph{domains}. Using this capability, we create a ``real-time'' domain for the victim task and a ``best-effort'' domain for the attackers. We then assign the BOOM core to the real-time domain and the two other Rocket cores, enhanced with the Mempress traffic generators, to the best-effort domain. As in Section~\ref{sec:attack}, Mempress instances are configured to generate overwhelming traffic to cache bank 0, to simulate the worst-case. 
For the victim tasks, we use \textit{BkPLLRead} (synthetic) and \textit{Disparity} (real-world), both configured to target cache bank 0. We vary the best-effort (attacker) domain's bandwidth budget from 640MB/s to 15.36GB/s, measuring the slowdown that each victim experiences normalized to the solo victim run (no attackers). The budget is set by programming the \textit{Regulation Period Register} to 400 cycles (400 ns in our setup), and increasing the best-effort domain's \textit{Access Budget Register} from 16 accesses (640MB/s) per-period to 384 accesses (15.36GB/s) per-period. Unless otherwise mentioned, all subsequent experiments use a regulation period of 400 cycles.

\begin{figure} [htp]
    \centering
    \includegraphics[width=0.5\textwidth]{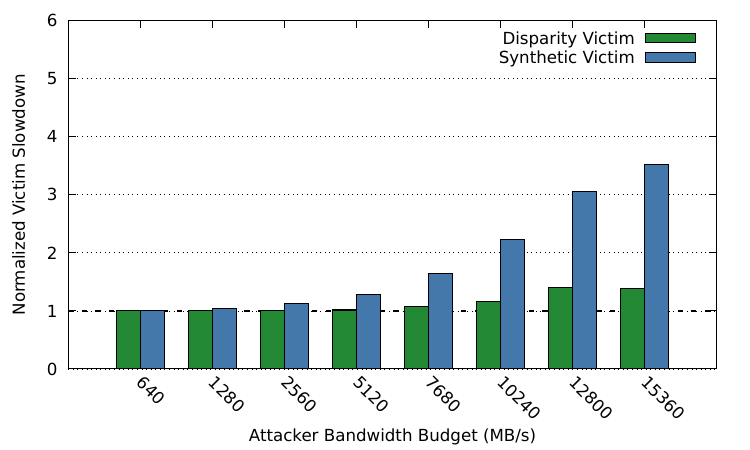}
    \caption{Impact of increasing attacker bandwidth budget on BkPLLRead and Disparity. The WSS of attackers is 64KB.}
    \label{fig:synthetic-budget-slowdown}
\end{figure}

Figure~\ref{fig:synthetic-budget-slowdown} shows the results. As we can see, up to an attacker budget of 1.28GB/s, the BkPLLRead victim experiences a 1.03$\times$ slowdown, which is small and  acceptable in may applications. Beyond this threshold, however, the victim’s execution begins to be impacted, increasing to 1.12$\times$ at a budget of 2.56GB/s and growing to 3.52$\times$ at 15.36GB/s. 
Note that 15.36GB/s is bigger than the observed peak cache memory bandwidth of the attackers, which is effectively identical to not using the bandwidth regulation at all.
When we repeat the experiment with \textit{Disparity} as the victim, the performance degrades at a slower rate, peaking only at 1.39$\times$ slowdown when the attackers are allotted their full bandwidth. This is because Disparity, unlike BkPLLRead, generates fewer accesses to the shared cache that are more evenly distributed among the cache banks. In other words, Disparity is not the worst-case and its performance impact from the cache bank attack will be upper bounded by that of the BkPLLRead victim. Since Disparity has the highest measured bandwidth of the SD-VBS workloads (see section~\ref{sec:sdvbs-profile}), all other workloads are similarly upper bounded.

The key takeaways are (1) cache bandwidth regulation can effectively regulate the attackers to provide worst-case slowdown guarantees for the victim; (2) the regulation budget should be set based on the worst-case scenario when both the attackers and the victim target one single cache bank. 

\subsection{All-Bank vs. Per-Bank Regulation} \label{sec:globalvper}
In the following experiments, we evaluate how different bandwidth regulation methods impact the performance of the victim and the attackers. 

We first show the impact of the all-bank and per-bank regulation methods in providing isolation guarantees to the victim task in the presence of the co-running cache-bank DoS attacks. 

The experiment setup is the same as before: the victim (BkPLLRead) runs on the real-time domain and the attackers run on the best-effort domain, which is regulated. The regulation budget of the best-effort domain is configured at 1.28GB/s (found to be the maximum allowable budget in the previous experiment)
in both all-bank and per-bank regulation methods. Note that under per-bank regulation, each bank receives the budget of 1.28GB/s.

\begin{figure} [htp]
    \centering
    \includegraphics[width=0.5\textwidth]{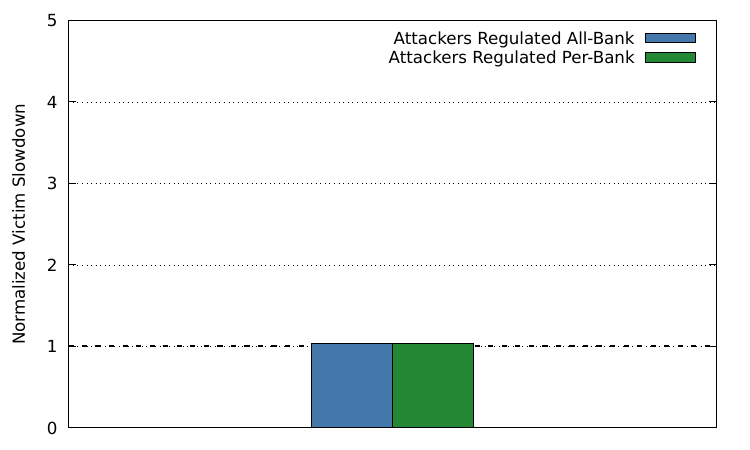}
    \caption{Normalized slowdown of the victim when a 1.28GB/s regulation budget is applied to throttle write attackers under both all-bank and per-bank regulations.}
    \label{fig:regattackers}
\end{figure}

Figure~\ref{fig:regattackers} shows the results. When running without regulation, we see the same 3.52$\times$ slowdown of the victim as was shown in the previous section. On the other hand, for both regulation schemes, the victim sees only a 1.03$\times$ slowdown with regulated attackers. This is because in both per-bank and all-bank regulation methods, only one cache bank is stressed by the attacks and the requests to the same bank are charged equally in both regulation methods. 

The results demonstrate that all-bank and per-bank bandwidth regulation methods are identical in protecting the victim in the worst-case (i.e., the cache bank DoS attackers are running on the best-effort domain). However, they will have very different effects in non worst-case scenarios as we will show in the following.

Next, we evaluate the throughput impact of the regulation methods on the regulated cores. 
For this, we use the \textit{Bandwidth} workload from~\cite{valsan2016taming} as described in section~\ref{sec:setup}. We configure the workload to perform read accesses with a WSS of 128KB (4$\times$ the L1 size). The workload is pinned to the system’s BOOM core. We measure the slowdown normalized to a non-regulated run of the workload. This experiment is performed on four different LLC designs as follows:

\begin{enumerate}
    \item All-bank regulation with two LLC banks.
    \item Per-bank regulation with two LLC banks.
    \item All-bank regulation with four LLC banks.
    \item Per-bank regulation with four LLC banks.
\end{enumerate}

\begin{figure} [htp]
    \centering
    \includegraphics[width=0.5\textwidth]{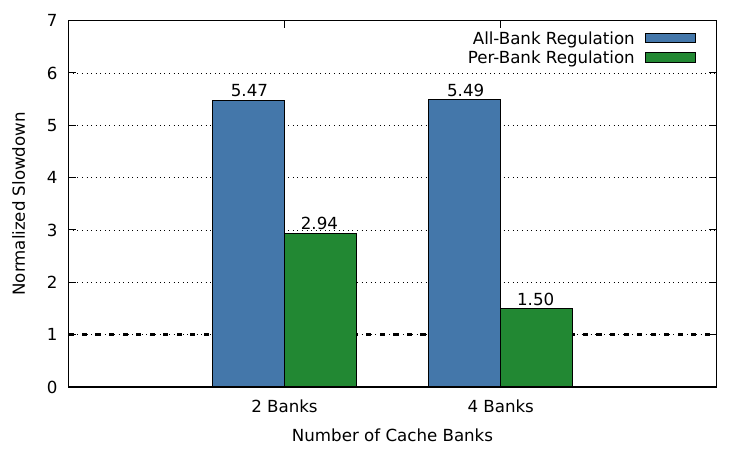}
    \caption{Normalized slowdown of \textit{Bandwidth} using per-bank and all-bank regulations on two and four bank cache configurations. Regulation budget is 1.28GB/s. Workload is pinned to BOOM core.}
    \label{fig:synth-compare}
\end{figure}

Figure~\ref{fig:synth-compare} shows the results. For the synthetic best effort task, regulating the entire cache as one unit (all-bank) results in performance degradation of 5.47$\times$ in the two-bank case and 5.49$\times$ in the four-bank case. In contrast, per-bank regulation sees a 2.94$\times$ and a 1.50$\times$ degradation in the two and four-bank cases, respectively. To directly compare, per-bank sees a 1.86$\times$ and 3.66$\times$ improvement over all-bank in the respective cases. Recall that this improvement is due to per-bank regulation allotting each bank a budget of 1.28GB/s. It should be noted that one would expect a 2$\times$ difference in the two bank cases and a 4$\times$ difference in the four bank cases. Of course, our prototype has some inefficiencies, however we deem these acceptable as the benefits of per-bank regulation are still clear.

From this synthetic experiment, we draw two major conclusions. First, per-bank regulation demonstrates a clear improvement in best-effort task throughput compared to the overly pessimistic all-bank regulation. This throughput improvement is achieved all while guaranteeing the same temporal isolation of victim tasks. Second, performance benefits of our per-bank implementation scale effectively as the number of banks increases.

\subsection{Impact of Per-Bank Regulation on Real-World  Applications} \label{real-world}

\begin{figure*}[htp]
  \centering
  \begin{subfigure}{0.49\textwidth}
    \includegraphics[width=\textwidth]{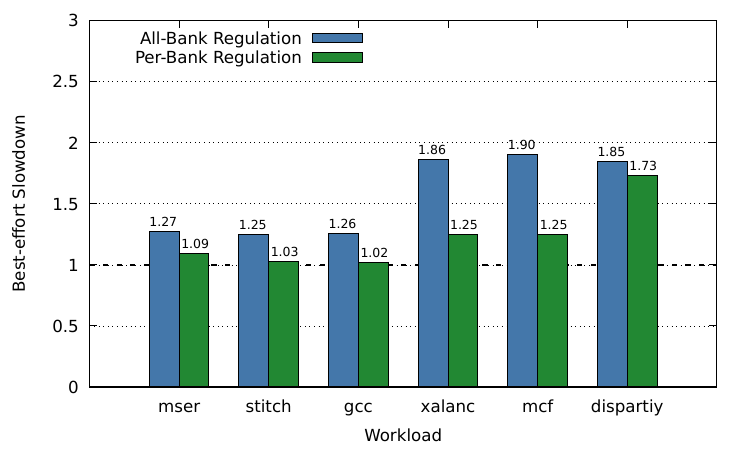}
    \caption{ 2 Banks }
    \label{fig:disparity-solo}
  \end{subfigure}
  \begin{subfigure}{0.49\textwidth}
    \includegraphics[width=\textwidth]{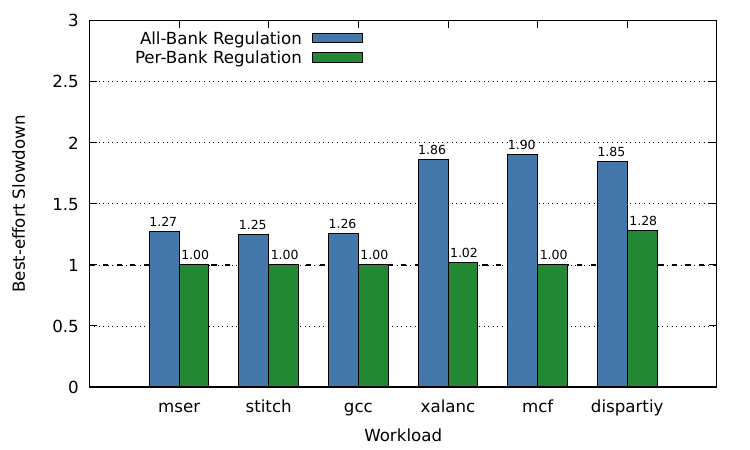}
    \caption{ 4 Banks }
    \label{fig:mser-solo}
  \end{subfigure}
  \caption{ Comparison of all-bank and per-bank regulation when running real-world workloads. Each workload is run 
  with a regulation budget of 1.28GB/s. Slowdown is in comparison to the unregulated run. }
  \label{fig:solo-real-all}
\end{figure*}

The \textit{Bandwidth} workload is a synthetic workload, not representative of real-world applications. We further evaluate the benefits of per-bank regulation over all-bank regulation using a set of benchmarks from SD-VBS~\cite{sdvbs} and SPEC2017~\cite{spec2017}. Specifically, we select \textit{Disparity}, \textit{MSER} and \textit{Stitch} from SD-VBS (all \textit{CIF} input) and \textit{gcc}, \textit{xalanc} and \textit{mcf} from SPEC2017 (\textit{ref} input). The SD-VBS benchmarks are chosen because they are relatively LLC bandwidth intensive workloads (see Section~\ref{sec:sdvbs-profile}). Likewise, the SPEC2017 benchmarks are chosen as they are relatively cache sensitive and have fast simulation run times.

For these experiments, we configure a system with one BOOM core to avoid any cross-core interference. We set a regulation budget of 1.28GB/s and measure each workload's performance, computing the slowdown normalized to an unregulated run of the workload. As was done in section~\ref{sec:globalvper}, we evaluate using both two bank and four bank cache designs.

Figure~\ref{fig:solo-real-all} shows the results for our selected workloads. In general, we see improvement when using per-bank regulation over all-bank regulation. Moreover, performance scales well from two to four banks, such as in \textit{Disparity} which suffers less slowdown as more cache banks are used.
Specifically, in the two-bank case, \textit{Disparity} sees a 1.85$\times$ and 1.73$\times$ slowdown under all-bank and per-bank regulation respectively. When the cache is configured with four banks, all-bank regulation creates the same 1.85$\times$ slowdown, while per-bank regulation has only a 1.28$\times$ slowdown. Thus, Disparity is an excellent example of the improvement of per-bank over all-bank regulation and the performance scalability as the number of banks increases. The results for \textit{MSER, Stitch} and \textit{gcc} also show similar improvement when going from an all-bank to per-bank regulation scheme.

These experiments clearly demonstrate that, as with the synthetic case, real-world workloads see noticeable improvement when using per-bank regulation over all-bank regulation. Again, it must be emphasized that this improvement is accompanied by per-bank regulation's guarantee of isolating victim tasks to the same degree as all-bank regulation. Overall, this highlights the superiority of per-bank regulation over all-bank.

\subsection{Software vs. Hardware Bandwidth Regulation}
In this experiment, we compare the software-based cache bandwidth regulation method proposed in~\cite{bechtel2023attack} with our hardware-based bandwidth regulation. 

Ideally, we would like to implement the software-based regulation approach directly on experimental platform. However, because we leverage Mempress traffic generators instead of using BOOM CPU cores for the attack, the software approach cannot be properly tested. Instead, we implement the software regulation approach on the BeagleV board, which is equipped with four RISC-V out-of-order cores (see Section~\ref{sec:realplatforms}) comparable to the BOOM core used in our testbed.

On the BeagleV platform, we observe up to 76$\times$ slowdown of the best-effort (attacker) tasks (throttled at 100MB/s) for the software regulation method to protect the victim. These results are in-line with the up to 300$\times$ slowdown reported in ~\cite{bechtel2023attack} on the Pi 4 platform. 
While using the hardware-based regulation methods in our FireSim setup, the worst-case slowdown of the attackers is up to 5.49$\times$ for all-bank regulation, and up to 2.94$\times$ slowdown for the per-bank regulation in 2-bank configuration. The slowdown is further reduced down 1.5$\times$ in the 4-bank configuration.

Note that the CPU performance of the BeagleV platform is on-par with that of our simulation setup. As such, we posit that the dramatic performance differences we observe come from the superior effectiveness of hardware-based regulation over the software-based one.

\subsection{Hardware Implementation Overhead}
To evaluate the cost-to-performance benefit of our per-bank design, we synthesize a quad-core BOOM SoC and perform power and area analysis. We run place and route from the Cadence Suite's Innovus tool, along with the Hammer~\cite{liew22hammer} VLSI flow scripts targeting the ASAP 7nm technology node~\cite{clark16asap7}, to characterize the overhead and compare to the all-bank implementation in~\cite{farshchi2020bru}.

\begin{table}[htpb]
	\centering
    \begin{adjustbox}{width=.48\textwidth}
	\begin{tabular}{|c||c|c|c|}
    	\hline
            Implementation & Regulation Unit (nm$^2$) & SoC (nm$^2$) & Percent \\
    	\hline
    	All-Bank\cite{farshchi2020bru} & 429 & 465305 & 0.09\%\\
    	\hline
    	Per-Bank (Ours)& 1372 & 466248 & 0.29\%\\
    	\hline
	\end{tabular}
    \end{adjustbox}
	\caption{Comparative area analysis of the two regulation unit implementations. Percent is the area consumed by the implementation from the total SoC area.}
	\label{tab:area}
\end{table}

Table~\ref{tab:area} shows the area utilization of the two configurations after place and route has been completed.
We find that the area overhead added in our per-bank design comes to 3.2$\times$ that of the all-bank implementation. However, it is still less than 0.3\% of the entire SoC area. 

\begin{table}[htpb]
	\centering
	\begin{tabular}{|c||c|c|}
    	\hline
            Design Under Test & Total Power (mW) & Percent \\
    	\hline
    	   SoC & 110 & \\
    	\hline
    	All-Bank\cite{farshchi2020bru} & 0.67 & 0.6\%\\
    	\hline
            Per-Bank (Ours)& 2.36 & 2.1\%\\
            \hline
	\end{tabular}
	\caption{Comparative power analysis of the two regulation unit implementations. Percent is the power consumed by the implementation from the total SoC power.}
	\label{tab:power}
\end{table}

Table~\ref{tab:power} shows the power analysis results. As shown, the per-bank design again consumes 3.5$\times$ that of the all-bank design, yet it is still only just over 2\% of the total power. 
It can be stated that the area and power overhead of our per-bank design is acceptable considering its significant performance benefits seen in previous sections.

\section{Related Work} \label{related}

In the real-time community, correctly estimating worst-case task execution timing is of paramount importance, yet it has been difficult to do so in multicore systems due to the vast and diverse set of shared hardware resources that can dramatically impact task execution timing. Microarchitectural DoS attacks on shared hardware resources are therefore important for the real-time community to study as they can shed light on the impacts on worst-case timing. 
Moscibroda et al. proposed the ``memory performance attack''~\cite{moscibroda2007memory}, which exploits the DRAM controller's FR-FCFS~\cite{rixner2000memory} scheduling policy to induce contention. Attacks on shared cache space~\cite{keramidas2006preventing}, bus bandwidth~\cite{woo2007analyzing}, shared cache MSHRs~\cite{valsan2016taming} and write-back buffers~\cite{bechtel2019dos}, shared GPU~\cite{yandrofski2022making}, and shared cache between the CPU and the integrated GPU~\cite{bechtel2022denial} have been explored. 
Most recently, bank contention on multi-bank shared caches has shown to be an effective DoS attack avenue~\cite{bechtel2023attack}, which we focus on in this work. 
Several studies have investigated the effects of these microarchitectural attacks in actual cyber-physical systems~\cite{li2024empirical,bechtel2024industry}.

Providing strong isolation in multicore has long been a topic of intense research over the past two decades. This includes various software and hardware mechanisms to manage the shared resources~\cite{mancuso2013rtas,kim2013coordinated,ye2014coloris,kim2017attacking,roozkhosh2020potential,yun2013memguard,ewarp20,farshchi2018deterministic,ali2019rt,xu2019holistic,saeed2022utilization}. Broadly, these resource management studies fall into two categories: space partitioning and bandwidth throttling. 
Cache space partitioning has been extensively studied in the real-time community to prevent unwanted cache-line evictions of high-priority real-time tasks by lower priority tasks. Cache space partitioning can be realized in software, through page coloring~\cite{mittal2017survey}, or in hardware, such as the way-based partitioning capabilities found in Intel RDT~\cite{rdt} and ARM MPAM~\cite{mpam}. 

Memory bandwidth throttling is another extensively studied mechanism for isolation. Most software-based memory bandwidth throttling techniques utilize the CPU core's performance counters to monitor the bandwidth. Then the periodic timers and interrupts regulate the allowed bandwidth of the cores at fixed time intervals~\cite{yun2013memguard,saeed2022utilization}. MemPol~\cite{zuepke2023mempol} instead utilizes a dedicated real-time micro-controller unit (MCU) to asynchronously monitor and regulate the memory traffic through polling. This approach reduces the interrupt overhead at the expense of wasting the real-time MCU.  
Hardware-based bandwidth regulation can eliminate such software overhead and can be enforced at a very fine granularity (in cycles rather than in milliseconds). BRU~\cite{farshchi2020bru}, Intel RDT~\cite{rdt}, ARM MPAM~\cite{mpam} all provide memory bandwidth regulation capabilities in hardware. 

Until recently, cache bandwidth has received little attention as it was believed to be of less importance compared to cache space partitioning or memory bandwidth. However, a recent study demonstrated its implications in multi-bank caches within high-performance multicore architectures~\cite{bechtel2023attack}. The study proposed a software cache bandwidth throttling mechanism as a potential mitigation solution, but acknowledged the unacceptably high overhead of such a software implementation. In this work, we present a hardware solution to manage cache bandwidth in real-time systems. To the best of our knowledge, we are the first to present a hardware-based cache bandwidth regulation solution. More importantly, we are the first to propose a per-bank cache bandwidth regulation approach that can significantly improve average throughput on the regulated cores. 

\section{Conclusion} \label{conclude}
In this paper, we presented a per-bank cache bandwidth regulation approach to effectively and efficiently mitigate potential cache bank bandwidth contention. We make the observation that the contention occurs at the individual cache bank rather than at the interconnect (bus), therefore our key contribution is to apply a well-known bandwidth regulation mechanism at the cache bank level. We evaluate that this approach can effectively minimize the effect of worst-case cache bank contention while maximizing allowed cache bandwidth and guaranteeing the isolation. We implemented the proposed per-bank regulation solution in hardware by extending an open-source bandwidth regulator design. We demonstrated its effectiveness in providing isolation guarantees to critical real-time tasks in the presence of adversarial cache bank DoS attackers. Furthermore, we illustrated that our per-bank bandwidth regulation approach can significantly improve performance of throttled best-effort tasks without compromising isolation guarantees allotted to real-time tasks. Specifically, we achieved up to 3.66$\times$ throughput improvement over the baseline bank-oblivious bandwidth throttling approach. 
As future work, we plan to apply the proposed per-bank regulation approach to DRAM banks. 

\section*{Acknowledgments}\label{sec:acknowledge}
This research is supported in part by NSF grants CPS-2038923, CCF-2239020, CCF-2403013, and ACE, one of the seven centers in JUMP 2.0, a Semiconductor Research Corporation (SRC) program sponsored by DARPA. 

\bibliographystyle{IEEEtran.bst}
\bibliography{references}

\end{document}